\documentclass[useAMS,usenatbib]{mn2e}
\usepackage[dvips]{graphicx}
\title[Chemistry of dense clumps near moving Herbig-Haro objects]{Chemistry of dense clumps near moving Herbig-Haro objects}
\author[H.Christie et al.]{H.Christie$^{1}$\thanks{E-mail:
hc@star.ucl.ac.uk}, S.Viti$^{1}$, D.A.Williams$^{1}$, J. M. Girart$^{2}$, O. Morata$^{3}$\\
$^{1}$Department of Physics and Astronomy, UCL, Gower street. London WC1E 6BT, \\
$^{2}$Institut de Ci\`encies de l'Espai (CSIC-IEEC), Campus UAB,
Facultat de Ci\`encies, Torre C5 - parell 2, 08193 Bellaterra, Catalunya, Spain, \\
$^{3}$Institute of Astronomy \& Astrophysics, Academia Sinica, P.O. Box 23-141, Taipei 10617, Taiwan}

\begin{document}

\pagerange{\pageref{firstpage}--\pageref{lastpage}} \pubyear{2009}

\maketitle

\label{firstpage}

\begin{abstract} Localised regions of enhanced emission from
HCO$^{+}$, NH$_3$ and other species near Herbig-Haro objects (HHOs)
have been interpreted as arising in a photochemistry stimulated by the
HHO radiation on high density quiescent clumps in molecular
clouds. Static models of this process have been successful in
accounting for the variety of molecular species arising ahead of the jet; however recent observations show that the enhanced molecular emission is widespread along the jet as well as ahead. Hence, a realistic model must take into account the movement of the radiation field past the clump. It was previously unclear as to whether the short interaction time between the clump and the HHO in a moving source model would allow molecules such as HCO$^{+}$ to reach high enough levels, and to survive for long enough to be observed. In this work
we model a moving radiation source that
approaches and passes a clump. The chemical picture is qualitatively unchanged by the addition of the moving source, strengthening the idea that enhancements are due to evaporation of molecules from dust grains. In addition, in the case of several molecules, the enhanced
emission regions are longer-lived. Some photochemically-induced
species, including methanol, are expected to maintain high abundances for
$\sim$10$^{4}$ years.

\end{abstract}


\section{Introduction}

Herbig-Haro objects are knots of optical emission, produced when the
jet from a young star collides with the ambient interstellar material
to produce a shock front (Falle \& Raga 1993). These
Herbig-Haro objects (HHOs) are strong line emitters at optical and
other wavelengths and are often seen along a protostellar outflow as
a series of bow shocks moving away from the star. A number of
observational surveys have detected localised regions of enhanced
emission in several molecules, among them NH$_3$ and HCO$^{+}$, just
ahead of Herbig-Haro objects (Girart et al. 1994; Girart, Estalella \&
Ho 1998, Torrelles et al. 1992). The regions appear chemically similar to each other and are
quiescent and cool. Therefore they are probably dynamically unaffected by
the jet. Girart et al. (1994) suggested that emission may be from molecules
in icy mantles on dust being released by UV radiation from the
Herbig-Haro object. Taylor \& Williams (1996) supported this theory
with a simple chemical model which reproduced the abundances inferred
from observations of these quiescent regions. A more complex chemical
model was then investigated by Viti \& Williams (1999) and used to
predict other molecular species expected to show enhanced emission under the
same conditions. Many of these, including CH$_3$OH, H$_2$CO and
SO$_2$, were later observed both in clumps ahead of HH2 (Girart et
al. 2002) and near to five other Herbig-Haro objects (Viti et al. 2006). 

The model used by Viti \& Williams (1999), Girart et
al. (2002) and Viti et al. (2003) described the particular
photochemistry that is produced when radiation from these HHOs
impinges on clumps of gas, located ahead of the bow shock, in which ices have returned to the gas
phase. In particular, the HHO (and hence the source of UV radiation) was assumed to be static.
However, recent observations of the object HH43 (Morata et al. in preparation) reveal 
the presence of several molecular species {\it along} the jet, where at least three HH objects  are present (see Figure 1). From the H$^{13}$CO$^+$ emission it is clear that the emission is in clumps or small filaments along the outflow and 
that they are quiescent (as they show narrow line widths). 
HCO$^+$ and CS emission also show such distinct clumps, and, consistent with previous observations of clumps ahead of HH objects, 
there seems to be a stronger contrast in intensity between clumps and elsewhere nearby in HCO$^+$ than in CS (consistent also with the 
fact that CS should trace larger scale gas).
These observations seem to indicate that quiescent clumps, chemically (but not dynamically) affected by the HH object, are present 
{\it along} the jet and not only ahead as previous surveys indicated. 
While previous modelling was successful at 
providing an explanation for the chemical enhancement ahead of the HHO,
what is now required is a model that can explain different degrees of chemical quiescent enhancements along the jet. The model needs therefore
to be dynamical in order to take into account the movement of the shock
front, typically travelling at a few hundred kms$^{-1}$ through a
molecular cloud. Raga \& Williams (2000) investigated, using a simple
chemistry, the effect of a moving field on the expected morphology of
the emission but the full consequences on the chemistry of allowing
the radiation source to move have not been explored. 

In this scenario, the HHO (the source of the radiation driving the 
photochemistry) approaches a clump and then passes it, so that the 
radiation intensity rises to a peak value and then decays. In this work we 
explore how the chemistry induced in the clump differs from that in the 
static case previously discussed, and consider the sensitivity of the 
chemistry to assumed geometry and physical conditions. Figure 1 suggests 
that in HH43 a clump may be affected by the passage of more than one HHO. 
However, in this work we examine the photochemistry induced by a single 
HHO passage.

Figure 1 shows the HH 38-43-64 system in emission lines of several
molecular species. This system of HHOs is initiated by a source, HH 43 MMS1,
indicated by a cross in the figure. The source appears to have initiated
several HHO events, but in our treatment we consider only a single event.
The figure shows clearly that emitting molecules are distributed along the
line of the jet, and are not confined to discrete objects in front of the jet
head.

\begin{figure*}
\vspace{30pt}
\includegraphics[width=4.5cm,angle=270]{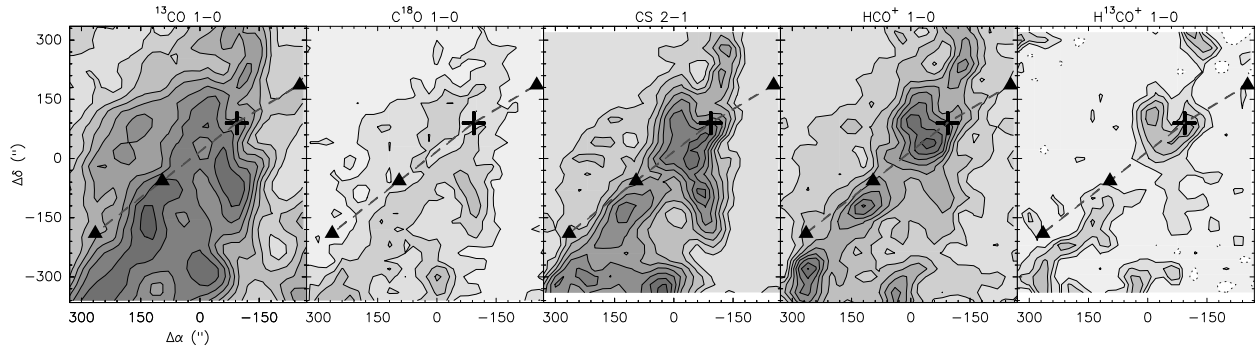}
\caption{Integrated emission of some molecular line transitions in the
6.2-7.2
kms$^{-1}$ $v_{\rm LSR}$ range, where emission (especially in
HCO$^+$)
follows HH 38-43-64 outflow. The molecular line transition is shown
on
the top of each panel.  For the C$^{18}$O and H$^{13}$CO$^+$ panels,
the contour levels are from 25\% to 95\% of the peak intensity in
steps of 20\%. For the other panels the contour levels are from 25\%
to 95\% of the peak intensity in steps of 10\%. The triangles show,
from left to right, the Herbig-Haro objects HH 38, HH 43 and HH 64.
The cross shows the position of HH 43 MMS 1, where the powering
source
of the HH system is located (Stanke, McCaughrean \& Zinnecker 2000).
Note that the two well defined clumps in HCO$^+$ ahead and south of
HH
43 and HH38 have narrow lines ($\Delta v\simeq 0.7$ kms$^{-1}$),
which suggests that they are dynamically quiescent.}

\end{figure*}

\section[]{The Model}

The basic model used is that of Viti \& Williams (1999) and Viti et
al. (2002) which runs in two phases, the first simulating the collapse
of molecular cloud gas from a fairly diffuse state to a clump of
uniform high density, and the second the illumination of the clump by
a static radiation source. 

In the model, the clump is treated as a one-dimensional slab of fixed temperature, increasing in visual extinction throughout up to a maximum value representing the clump centre. Abundances are calculated for 10 depth points through the slab. We assume the clump to be spherically symmetric and the radiation field to be isotropic at the clump surface so that the 1D slab is able to represent the whole clump. Using classic rate equations and the abundances in the previous time step, abundances of species are calculated at each depth point and for each time step (the gap between time steps is varied according to how quickly abundances are likely to be changing). This way the chemistry of the whole clump is tracked for the duration of the model run. Self-shielding of molecular hydrogen and CO is taken into account so that photodissociation of these species depends on the abundances in outer depth points. 

The model follows the
chemical evolution of 170 species including 1858 separate reactions
for the 10 depth points of increasing visual extinction (A$_v$).
Reaction rates are taken from the UMIST database (Woodall et
al. 2007).

During phase I the clump undergoes a modified free-fall collapse during which molecules are allowed to freeze-out or deplete onto dust grains, where they undergo hydrogenation as far as possible. Ions are neutralized on hitting the grain surface and reaction rates on the surfaces take into account the small negative charge on the dust grains resulting in a slightly higher rate for the positive ions. The radiation field in phase I is fixed at 1 Habing to represent the ambient interstellar field.

\rm{Depletion of species from the gas phase as they are frozen out onto
grains is controlled within the model by effectively altering the
grain surface area available for gas species to freeze-out. The
freeze-out fraction of CO at the end of phase I was set to around 20\%
for models with a final clump density of 10$^{4}$ cm$^{-3}$ (regardless
of the initial clump density), around 50\% for final densities of
10$^{5}$ cm$^{-3}$ and 70\% for final densities of
10$^{6}$ cm$^{-3}$. These values are consistent with
observational depletion studies of isolated dark clouds, where denser
objects show a higher degree of freeze-out of CO (eg. Redman et
al. 2002). All models assumed that 5\% of the CO freezing out onto
the grains was converted into methanol.

\par The initial number density of hydrogen
nuclei in the clump (before collapse) is assumed to be 10$^{3}$ cm$^{-3}$. It is also
assumed that, initially, only carbon is ionised and half of the
hydrogen is in its molecular form. Nitrogen, oxygen, magnesium,
sulphur and helium are all neutral and atomic. As assumed by Viti \& Williams
(1999), the first phase continues until a specified number density is
reached. This number density is observationally
determined to be in the range 10$^{4}$ to 10$^{5}$ cm$^{-3}$
(e.g. Girart et al. 2005; Whyatt et al. 2010). We have also
explored the effect of allowing somewhat higher densities in several
models (see Table 1). For simplicity, the clump is assumed to be of
uniform density at all times. 
\par The second phase commences immediately after the specified density is
attained, when it is assumed that the radiation from the HHO is
switched on. There is no collapse in this phase, the clump remains at a constant density (equal to that at the end of phase I) and a constant temperature throughout. The HH radiation field is again isotropic but stronger than the ambient field irradiating the clump in phase I. The chemistry in the second phase is then computed for a model time of about a million
years.

The radiation field of the HHO is represented in the model in
terms of multiples, G$_0$, of the mean interstellar radiation field
intensity. This is clearly an approximation, since the spectra of the
two fields are not the same. Unfortunately, all astrochemical codes
use the interstellar field as the basis of their treatments of
photo-processes, and to do otherwise would be a major project beyond
the scope of this work. While the approximation we have used
could give misleading results, the prediction of a rich characteristic
photochemistry in HHO-illuminated clumps (Viti \& Williams 1999) has
been confirmed by observations of a number of sources (Girart et
al. 2002, 2005, Viti et al. 2006) and by a very detailed study of HH2
(Girart et al. 2005).

In the adapted model, the moving source of this
radiation field (representing the HHO) passes the clump
on a straight path with a minimum distance to the edge of
the clump, here chosen to be 0.05 pc. This distance 
is similar to
values found by Whyatt et al. (2010) in an observational study of regions around 22
HHOs. For a source moving at 300 kms$^{-1}$, there will be
enough time for it to reach closest approach (at around 1000 years)
and move some distance away before phase II is terminated. There is a range of values observed for HHO
velocities from $\sim$100 to $\sim$1000 kms$^{-1}$. Faster HHOs have a
shorter interaction time. We have used 300
kms$^{-1}$ for most of the models considered, but have also explored
the effects of a higher velocity in models 16-18 (see Table 1). The
flux reaching the edge of the clump is time-dependent, changing with
the distance from source to clump and reaching a
specified peak value. Absorbing material along the line of sight to the clump is represented by an A$_v$ of 1 magnitude at the clump edge. This low extinction material will, in any case, have little
effect on the model outputs. Models with a static radiation source (kept at 0.05pc from the
clump throughout phase II) were run for comparison.

The main parameters affecting the model chemistry are the density of
the molecular cloud from which the clump forms in phase I and the
density of the clump at the end of phase I; the final clump radius
which, along with the density, will determine the visual extinction, A$_v$, of the clump;
the radiation peak strength of the field from the Herbig-Haro object at the
clump boundary; 
and the source velocity. The values of these parameters adopted
for each model are shown in Table 1. We also ran a few models to represent the interclump medium, i1, i2 and i3, with an initial density in phase I of 10$^{2}$ cm$^{-3}$ and a final density of 10$^{3}$ cm$^{-3}$. No freeze-out occurs in either phase for the interclump models and phase I is extended for a time after the final density is reached. 
The relative
radiation strengths, G$_0$, listed in Table 1 are up to a few tens;
these are the strengths found in previous modelling work (Viti \&
Williams 1999) to be necessary to create the rich variety of
photochemistry subsequently observed by Girart et al. (2002) and Viti
et al. (2006). Elemental abundances, adapted
from Sofia \& Meyer (2001), are listed in Table 2.

\begin{table}
  \caption{Model Input Parameters}
  \label{tab:initial}
  \begin{center}
    \leavevmode
    \begin{tabular}{ |p{1cm}|p{1cm}|p{1cm}|p{1cm}|p{1cm}|p{1cm}|} \hline 
   Model Number & Initial Cloud Density (cm$^{-3}$) & Final Clump Density (cm$^{-3}$) & Clump Radius (pc) & Radiation Field Strength (G$_0$) & Source velocity (kms$^{-1}$) \\ \hline
   i1 & 10$^{2}$ & 10$^{3}$ & 3 & 5 & 300\\
   i2 & 10$^{2}$ & 10$^{3}$ & 3 & 20 & 300\\
   i3 & 10$^{2}$ & 10$^{3}$ & 3 & 30 & 300\\
   1 & 10$^{3}$ & 10$^{4}$ & 0.3 & 5 & 300\\
   2 & 10$^{3}$ & 10$^{4}$ & 0.3 & 20 & 300\\
   3 & 10$^{3}$ & 10$^{4}$ & 0.3 & 30 & 300\\
   4 & 10$^{3}$ & 10$^{5}$ & 0.03 & 5 & 300\\
   5 & 10$^{3}$ & 10$^{5}$ & 0.03 & 20 & 300\\
   6 & 10$^{3}$ & 10$^{5}$ & 0.03 & 30 & 300\\
   7 & 10$^{4}$ & 10$^{5}$ & 0.03 & 5 & 300\\
   8 & 10$^{4}$ & 10$^{5}$ & 0.03 & 20 & 300\\
   9 & 10$^{4}$ & 10$^{5}$ & 0.03 & 30 & 300\\
   10 & 10$^{4}$ & 10$^{6}$ & 0.003 & 5 & 300\\
   11 & 10$^{4}$ & 10$^{6}$ & 0.003 & 20 & 300\\
   12 & 10$^{4}$ & 10$^{6}$ & 0.003 & 30 & 300\\ 
   13 & 10$^{3}$ & 10$^{4}$ & 0.3 & 50 & 300\\
   14 & 10$^{3}$ & 10$^{5}$ & 0.03 & 50 & 300\\
   15 & 10$^{4}$ & 10$^{6}$ & 0.03 & 50 & 300\\
   16 & 10$^{3}$ & 10$^{5}$ & 0.03 & 5 & 1000\\
   17 & 10$^{3}$ & 10$^{5}$ & 0.03 & 20 & 1000\\
   18 & 10$^{3}$ & 10$^{5}$ & 0.03 & 30 & 1000\\ \hline
    \end{tabular}
  \end{center}
\end{table}

\begin{table}
  \caption{Initial elemental abundances (from Sofia \& Meyer 2001)}
  \label{tab:initial}
  \begin{center}
    \leavevmode
    \begin{tabular}{ |p{3cm}|p{2cm}|} \hline 
   Element & Initial abundance (X(N)/X(H)) \\ \hline
   Helium & 0.075\\
   Carbon & 1.79$\times$10$^{-4}$\\
   Oxygen & 4.45$\times$10$^{-4}$\\
   Nitrogen & 8.52$\times$10$^{-5}$\\
   Sulphur & 1.43$\times$10$^{-6}$\\
   Magnesium & 5.12$\times$10$^{-6}$\\ \hline
    \end{tabular}
  \end{center}
\end{table}

Column densities are calculated by summing fractional abundances at
all 10 depth points for each species, taking into account the value of A$_v$ at
each point.

\section[]{Results} 
Our results are summarized in Figures 2 and 3.
While we find that the fractional abundances of a given
species in the gas surrounding an Herbig-Haro object are clearly
different between the `moving' and `static' cases, it is clear that the theory proposed by Viti \& Williams
is still valid, even if the radiation source is moving with respect to the clumps.  In fact, some species
seem to survive longer if a moving source is included.
Figure 2 plots column
densities of four selected species throughout phase II. Both the
moving and static source models are shown as well as column densities for the interclump medium, also under the influence of moving and static fields. The four species included in Figure 2
were selected  because their abundances were particularly altered by a moving source or because
they are particularly important
observationally. Discrepancies were
largest for lower A$_v$ (smaller or more diffuse clumps) but similar
for radiation field strengths in the range 5-50 G$_0$, so
the fact that the radiation source is moving is important,
regardless of field strength. Table 3 compares the effects of moving and static sources 
for several observationally important species.
Species whose abundances differ at least three orders of magnitude
at any point in phase II between the moving and static cases are
considered very strongly affected by the presence of the moving
field. Where abundances differ at least two orders of magnitude but less than three, species
are considered strongly affected. If abundances differ at least one
order of magnitude and less than two, species are considered weakly affected and where
abundances differ less than one order of magnitude between the two cases
species are considered unaffected. Table 3 refers to model 5 with
A$_v$ of 5 magnitudes. Arrows indicate the way in which the moving source affects the abundance
of each species both at early times and later, during the passage of the HHO.  

\begin{figure*}
\centering
\vspace{30pt}
\includegraphics[width=14cm,height=17cm,angle=270]{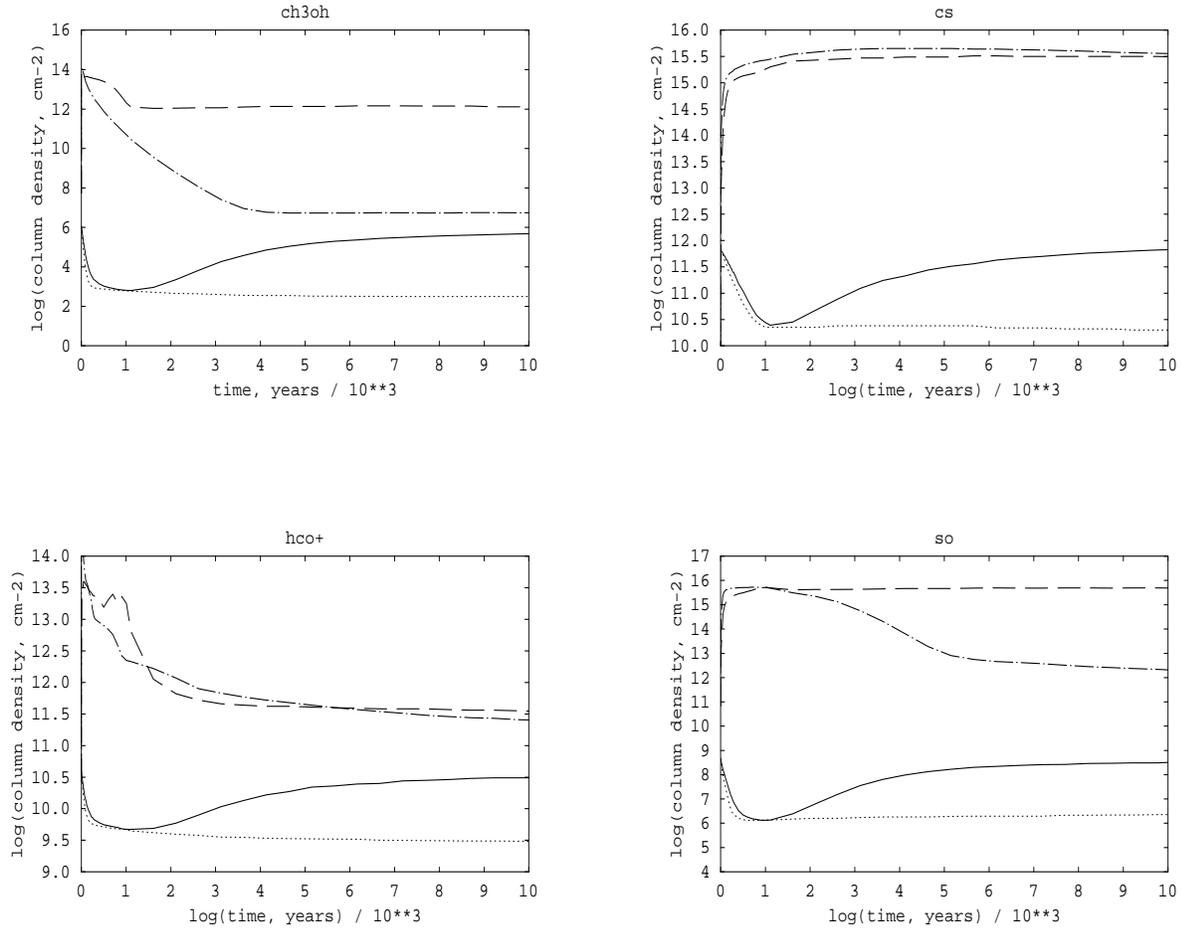}
\caption{Column density (cm$^{-2}$) versus time (years). The solid black line represents the interclump medium, A$_v$ $\sim$2 mags, irradiated by a moving field of 30 G$_0$ (model i3), dashes - a clump at 10$^{5}$ cm$^{-3}$, A$_v$ $\sim$5 mags, irradiated by a moving field of 30 G$_0$ (model 6), dots - the interclump medium, A$_v$ $\sim$2 mags, irradiated by a static field of 30 G$_0$ (model i3 with a static field) and dots and dashes - a clump at 10$^{5}$ cm$^{-3}$, A$_v$ $\sim$5 mags, irradiated by a static field of 30 G$_0$ (model 6 with a static field). In the moving case the radiation field source is at its closest point at around 1000 years.}
\end{figure*}

\begin{table}
  \caption{Comparing the effects of moving and static sources for model 5 with the clump at an A$_v$ of 5 magnitudes. E denotes early times, L late times (after around 300 years). Up arrows indicate molecules that increase in abundance with a moving source rather than static, right arrows those that do not change and down arrows those that decrease in abundance.}
  \label{tab:initial}
  \begin{center}
    \leavevmode
\tiny    \begin{tabular}{|p{0.6cm}|p{0.1cm}|p{0.1cm}|p{0.6cm}|p{0.1cm}|p{0.1cm}|p{0.6cm}|p{0.1cm}|p{0.1cm}|p{0.6cm}|p{0.1cm}|p{0.1cm}|} \hline 
\multicolumn{3}{|p{1cm}|}{\tiny{Very Strongly Affected}} & 
\multicolumn{3}{|p{1cm}|}{\tiny{Strongly Affected}} & 
\multicolumn{3}{|p{1cm}|}{\tiny{Weakly Affected}} & 
\multicolumn{3}{|p{1cm}|}{\tiny{Not Affected}} \\ \hline
Mol & E & L & Mol & E & L & Mol & E & L & Mol & E & L\\ \hline
\tiny{CH$_3$OH} & $\rightarrow$ & $\uparrow$ & \tiny{OCS} & $\rightarrow$ & $\uparrow$ & \tiny{CN} & $\rightarrow$ & $\downarrow$ & \tiny{H$_2$CO} & $\downarrow$ & $\downarrow$ \\
\tiny{NH$_3$} & $\rightarrow$ & $\uparrow$ & \tiny{HC$_3$N} & $\uparrow$ & $\uparrow$ & \tiny{HCN} & $\uparrow$ & $\rightarrow$ & \tiny{HCO} & $\downarrow$ & $\uparrow$ \\
\tiny{SO$_2$} & $\downarrow$ & $\uparrow$ & \tiny{C$^{+}$} & $\downarrow$ & $\downarrow$ &\tiny{C$_3$H$_5$$^{+}$} &$\uparrow$ &$\rightarrow$ & \tiny{CS}&$\rightarrow$ &$\downarrow$  \\
\tiny{SO} & $\downarrow$ & $\uparrow$ & \tiny{OCN} & $\rightarrow$ & $\uparrow$ & \tiny{H$_2$CN} & $\rightarrow$ &$\downarrow$ & \tiny{HCO$_+$} &$\uparrow$ &$\downarrow$  \\
\tiny{H$_2$S} & $\rightarrow$ & $\uparrow$ & \tiny{NS} & $\rightarrow$ &$\uparrow$ & \tiny{HNC} &$\rightarrow$ &$\downarrow$ & & &  \\
 & & & \tiny{CH$_3$CN} &$\rightarrow$ &$\uparrow$ & \tiny{HCS$^{+}$} &$\uparrow$ &$\downarrow$ & & &  \\
 & & & \tiny{C$_3$H$_4$} &$\uparrow$ &$\rightarrow$ & \tiny{H$_2$CS} &$\downarrow$ &$\uparrow$ & & &  \\
 & & & \tiny{CO} &$\downarrow$ &$\rightarrow$ &\tiny{NO$^{+}$} &$\rightarrow$ &$\uparrow$ & & &  \\
 & & & \tiny{C$_2$H} &$\uparrow$ &$\uparrow$ & \tiny{C} &$\downarrow$ &$\downarrow$ & & &  \\\hline
    \end{tabular}
  \end{center}
\end{table}
In general, including a moving source implies that the radiation
field decays quickly enough that the chemistry is being driven more
slowly. This way the specially created species arising in the
photochemistry survive for longer. For most of the strongly affected
species, with a moving radiation source, column densities may remain
high at least up to 30,000 years  (and possibly much longer)
after the passage of the source. Table 4 illustrates this
point: here we list the length of time selected species survive for
both the static and
moving source cases.   

\begin{table}
  \caption{Timescales of abundance enhancements - Model 5. Timescale defined as the time taken for column density to drop below 10$^{12}$ cm$^{-2}$ or to stop falling.}
  \label{tab:initial}
  \begin{center}
    \leavevmode
    \begin{tabular}{|p{1.5cm}|p{1.5cm}|p{1.5cm}|} \hline 
   Molecule & Timescale (Moving) & Timescale (Non-moving) \\ \hline
   CH$_3$OH & 10$^{5}$ yrs & 10$^{3}$ yrs\\
   NH$_3$ & 10$^{5}$ yrs & 10$^{4}$ yrs\\
   SO & 10$^{6}$ yrs & 5$\times$10$^{5}$ yrs  \\
   HCO$^{+}$ & 5$\times$10$^{3}$ yrs & 5$\times$10$^{3}$ yrs\\ 
   CN & 10$^{5}$ yrs & 10$^{5}$ yrs\\
   HCN & 10$^{6}$ yrs &  5$\times$10$^{5}$ yrs\\
   CS & 5$\times$10$^{6}$ yrs &  5$\times$10$^{5}$ yrs\\
   OCS & 5$\times$10$^{6}$ yrs &  10$^{3}$ yrs\\
   CO & 10$^{6}$ yrs & 10$^{5}$ yrs\\
   NS & 10$^{6}$ yrs & 10$^{6}$ yrs\\
   H$_2$CO & 10$^{6}$ yrs & 5$\times$10$^{5}$ yrs\\
   H$_2$S & 5$\times$10$^{4}$ yrs & 5$\times$10$^{3}$ yrs \\
   H$_2$CS & 5$\times$10$^{6}$ yrs & 5$\times$10$^{5}$ yrs\\\hline
    \end{tabular}
  \end{center}
\end{table}

Most species chosen for the study are first enhanced and later
destroyed by the radiation field and hence show similar behaviour
under the influence of a moving source. Their abundances are marginally lower
than for the static field for up to a few hundred years (in which time
the HHO moves only a very small distance) but later, as
species begin to be destroyed by the radiation, the abundances in the
moving case remain higher for longer. In some cases the chemistry
appears to be such that abundances may not return to their initial values for long periods. 
CH$_3$OH, H$_2$S and
NH$_3$ have higher abundances in the moving case than in the static case for the evolutionary time shown in Figure 2.
C and C$^{+}$ are enhanced by
the radiation field and abundances are lower at all times in the moving
case. HCO$^{+}$ and HCS$^{+}$ have higher abundances in the moving
case at very early times but drop more later and are lower than
for the static field at later times.

Models 13-15 were run with a 50 G$_0$ field. The changes in abundances
are similar to models with a weaker field although more extreme, with
elements such as CH$_3$OH and NH$_3$ decreasing in abundance at late
times. This is illustrated in Figure 3 which shows the effect (on
HCO$^{+}$) of altering A$_v$, field strength, shock velocity and
final clump density on the model output in phase II.

It appears counter intuitive that the density affects the abundance of HCO$^{+}$ in the opposite sense to the A$_{v}$. However, because the A$_{v}$ for the different density models is fixed (at around 6 magnitudes for the innermost depth point) the radiation field penetrates to the same extent in all models. The difference in abundance thus arises from differences in reaction rates due to the density of material for the main reactions forming and destroying HCO$^{+}$. These are, respectively, the recombination of CH and O and the recombination of HCO$^{+}$ with electrons.

The influence of the radiation source speed on the model output was
investigated in models 16-18 (again see Figure 3). It appears that a
faster moving source allows several important molecular species to
sustain, up to 30,000 years at least, higher column densities than in
the case of a 300 kms$^{-1}$ shock. Generally, the effects are seen
later than about 300 years.\\

\begin{figure*}
\vspace{10pt}
\hspace{-20pt}
\includegraphics[width=14cm,height=17cm,angle=270]{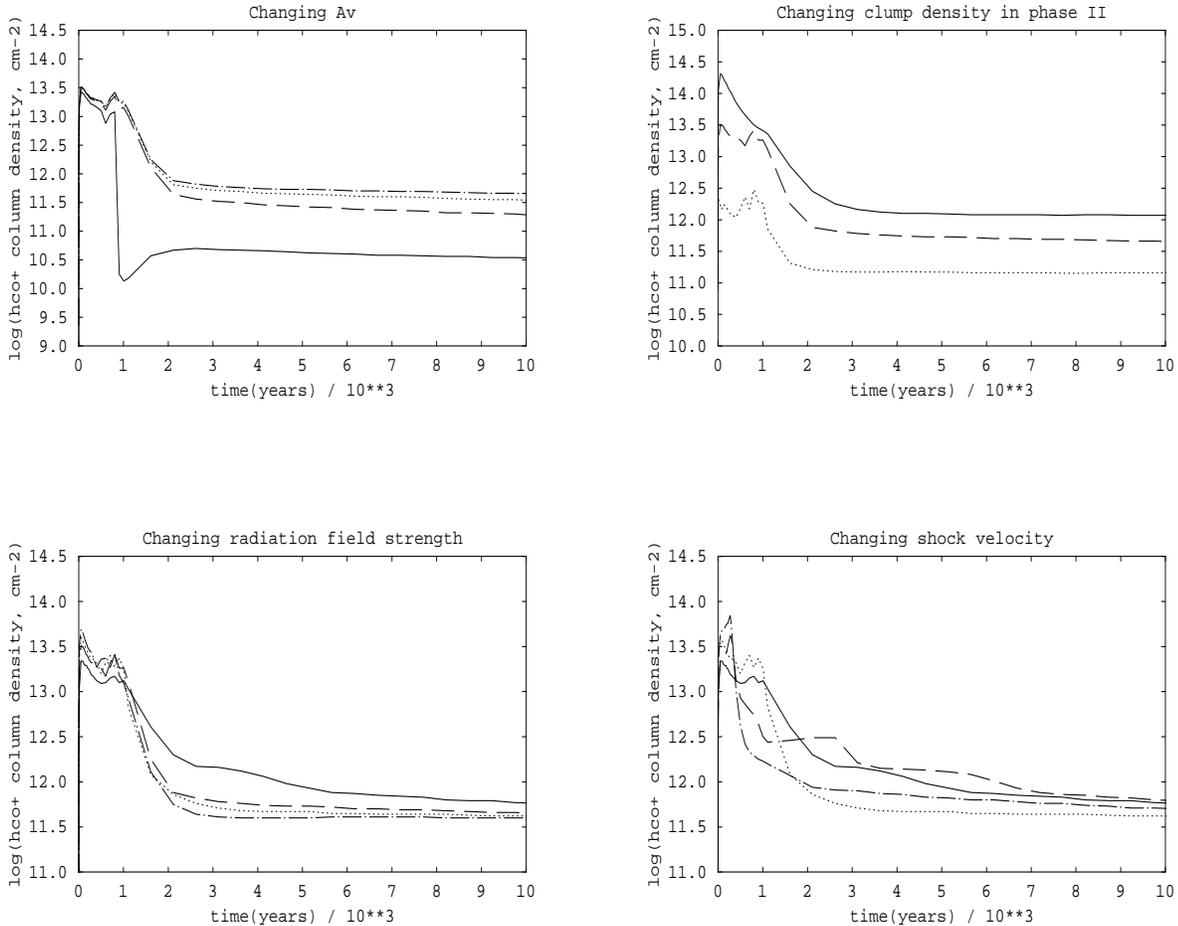}
\caption{For HCO$^{+}$. Top left plot - varying A$_v$: solid line represents a clump at 10$^{5}$ cm$^{-3}$ with a moving source of 20 G$_0$ (model 5) at 1 mag, dashes - 3 mags, dots - 4 mags, dots and dashes - 6 mags. Top right hand plot - varying clump density: solid line represents model 2 (10$^{4}$ cm$^{-3}$ with a moving source of 20 G$_0$), dashed line - model 5 (10$^{5}$ cm$^{-3}$), dotted line - model 11 (10$^{6}$ cm$^{-3}$). Lower left hand plot - varying radiation field strength: solid line represents model 4 (5 G$_{0}$), dashed line - model 5(20 G$_{0}$) , dotted line - model 6 (30 G$_{0}$)and single dots and dashes - model 14 (50 G$_{0}$) . Bottom right hand plot - varying shock velocity: solid line represents model 4 (5 G$_{0}$ at 300 kms$^{-1}$), dashed line - model 16 (5 G$_{0}$ at 1000 kms$^{-1}$), the dotted line - model 6 (20 G$_{0}$ at 1000 kms$^{-1}$) and dots and dashes - model 18 (30 G$_{0}$ at 1000 kms$^{-1}$). Apart from the top left hand plot A$_v$ $\sim$ 6}
\end{figure*}


\section{Conclusions}
Clumps containing enhanced molecular abundances are routinely observed near Herbig-Haro objects (HHOs) in low-mass star-forming regions. The
characteristic chemistry displayed by these clumps is consistent with a model in which the gas of
evaporated ices is subjected to a photochemistry driven by the nearby HHO. Previous models have
been successful in reproducing the variety in the observed chemical species; however, it was not obvious that they could explain the observed clumps along a jet which would be subject to a varying radiation field during the passage of the HHO. Moreover, the
chemical effects in previous models were transient on a short timescale so that clumps were required to be extremely young and short-lived, causing
some concern. In this work,
we investigate the effect of adapting the earlier models to include a
moving, rather than static, radiation source. The main
conclusions of the work are as follows:

\begin{itemize}

\item  Results from the moving source model confirm that it is still possible to reproduce the particular chemistry observed in clumps near to HHOs while allowing the radiation source to move rather than remain static relative to the clump. This supports the idea that emission is due to the evaporation of species frozen out onto dust grains in dense regions of a molecular cloud.

\item The new model enables several important molecules to maintain
detectable abundances for longer periods.

\item Species can be grouped
into roughly three categories displaying similar behaviour under the
influence of the moving source. Some (such as CH$_3$OH, NH$_3$ and
H$_2$S) have much higher abundances at all times with a moving source than
with a static source. Most species investigated (including SO$_2$, SO,
NO$^{+}$, HCO$^{+}$, HCN, CN, CS and OCS) are first formed and then
destroyed by reactions initiated by the radiation field and hence are less abundant with the
moving field up to about 1000 years (for a source moving at
300 kms$^{-1}$) and then more abundant (this is most apparent for lower A$_{v}$ clumps, hence is not obvious in figure 2 which plots those of higher A$_{v}$). C$^{+}$ and C abundances are
enhanced by the field and are lower with the moving field at all
times.  

\item Species with most noticeably increased abundances in the
`moving' case (as opposed to the `static' case) are CH$_3$OH, NH$_3$,
SO$_2$, SO and H$_2$S. 

\item The discrepancy between the moving and
static cases is greater for a faster moving, stronger source and clumps
of smaller size or lower density.  \end{itemize}

The results of this investigation support the idea that the observed chemistry ahead of Herbig-Haro objects is a result of species on the grains returning to the gas phase. The moving source allows the chemistry to persist for longer, helping to explain the large number of these clumps observed. 

\section*{Acknowledgements}

H.C. thanks Dermot Madden for help with coding and the STFC for funding.

\newpage
\clearpage
\label{lastpage}

\end{document}